\documentclass[superscriptaddress,endfloats, preprint]{revtex4-1}  % for review and submission
\usepackage{graphicx}% Include figure files

\def\figsizeA{16cm}
\def\figsizeB{16cm}
\def\figsizeC{18cm}

\begin{document}

\title{Topological Phase Transitions in Spatial Networks}

\author{Paul Balister$^\dagger$}
\affiliation{\footnotesize{Department of Mathematical Sciences, University of Memphis, Memphis TN 38152, USA}}
\author{Chaoming Song$^\dagger$}
\affiliation{\footnotesize{Department of Physics, University of Miami, Coral Gables, Florida 33142, USA}}
\author{Oliver Riordan$^\dagger$}
\affiliation{\footnotesize{Mathematical Institute, University of Oxford, Oxford OX2 6GG, UK.}}
\author{B\'ela Bollob\'as}
\affiliation{\footnotesize{Department of Pure Mathematics and Mathematical Statistics, Wilberforce Road, Cambridge
CB3 0WB, UK}}
\affiliation{\footnotesize{Department of Mathematical Sciences, University of Memphis, Memphis TN 38152, USA}}
\author{Albert-L\'aszl\'o Barab\'asi}
\affiliation{\footnotesize{Center for Complex Network Research,
Department of Physics, Biology and Computer Science, Northeastern
University, Boston, Massachusetts 02115, USA}}
\affiliation{\footnotesize{Department of Medicine, Brigham and
Women's Hospital, Harvard Medical School, Boston, Massachusetts
02115, USA}\\\footnotesize{$\dagger$ These authors contributed equally to this work}}

\begin{abstract}
Most social, technological and biological networks are embedded in a finite dimensional space, and the distance between two nodes influences the likelihood that they link to each other. Indeed, in social systems, the chance that two individuals know each other drops rapidly with the distance between them; in the cell, proteins predominantly interact with proteins in the same cellular compartment; in the brain, neurons mainly link to nearby neurons. Most modeling frameworks that aim to capture the empirically observed degree distributions tend to ignore these spatial constraints. In contrast, models that account for the role of the physical distance often predict bounded degree distributions, in disagreement with the empirical data. Here we address a long-standing gap in the spatial network literature by deriving several key network characteristics of spatial networks, from the analytical form of the degree distribution to path lengths and local clustering. The mathematically exact results predict the existence of two distinct phases, each governed by a different dynamical equation, with distinct testable predictions. We use empirical data to offer direct evidence for the practical relevance of each of these phases in real networks, helping better characterize the properties of spatial networks.
\end{abstract}

\def\rminj{\ell_j}
\def\rmini{\ell_i}
\def\rmax{L}
\def\r{\mathbf{r}}

\maketitle

In many real systems the likelihood that two nodes connect to each other decreases rapidly with the physical distance between them. For example, in \emph{social networks}, spatial proximity determines the likelihood that two individuals meet. Consequently, the probability $P(r)$ that two individuals, living at distance $r$ from each other, are connected by social media \cite{barthelemy2011spatial, liben2005geographic} and by mobile phone calls \cite{lambiotte2008geographical}, scientific collaboration \cite{pan2012world} and emails \cite{adamic2005search}, has been found to decay with $r$, often following $P(r) \sim r^{-\delta}$ with $\delta \approx 1$ \cite{liben2005geographic}. Similarly, in \emph{infrastructural networks} like the Internet and the power grid, the further apart two nodes are, the higher the installation cost of connecting them as well as the maintenance cost. Consequently, the likelihood that two Internet routers link to each other decays as $P(r)\sim r^{-1}$ \cite{yook2002modeling}. In the same vein, most transportation networks, such as highway and train networks, are planar, implying that their links predominantly connect nearby nodes \cite{masucci2009random}. While the airline network appears to have no such local constraints, measurements indicate that the probability $P(r)$ that two airports have a direct flight between them decays as $P(r) \sim r^{-\delta}$ with $\delta \approx 3$ for $r > 100$ km \cite{bianconi2009assessing}. The same phenomenon is known to occur in \emph{cellular networks}: The difficulty of subcellular molecular transport significantly limits the likelihood that two molecules that are not localized in the same cellular component can interact  \cite{srere2000macromolecular,sayyid2016importance}. And in the brain, notwithstanding the presence of very long axons, there is extensive evidence that neurons predominately link to neurons in their immediate vicinity \cite{bullmore2009complex}, with $P(r)$ being approximated by either a short range exponential $P(r) \sim \exp(-\lambda r)$ \cite{markov2013cortical, markov2013role, ercsey2013predictive}, or a long range decay $P(r) \sim r^{-\delta}$ \cite{vertes2012simple}.

Despite the well documented role of physical distance in real networks, the current modeling frameworks tend ignore the role of the space, for two reasons: First, while there are extensive numerical results unveiling the properties of spatial networks, most findings remain conjectural due to the lack of rigorous results \cite{barthelemy2011spatial}. Second, in many systems such as brain or cellular networks the location of the nodes is hard to measure. Here we focus on a much studied model that accounts in a minimal, yet realistic, fashion both for spatial effects and for the observed fat-tailed degree distribution of real networks \cite{yook2002modeling, manna2002modulated, xulvi2002evolving, boccaletti2006complex, poncela2008complex, gomez2004local}. Consider a $D$-dimensional map of linear dimension $L$. For $D = 2$ the model mimics a country or a continent, for $D=3$ it captures neuron connectivities in the brain or protein interactions in a cell. At each time step we place a new node $j$ on the map, its physical position $\r$ being chosen randomly following the density function $\rho(\r)$, which mimics for example the population density of a country or the protein density in a cellular compartment. Each new node connects to $m$ preceding nodes, chosen independently \cite{BRsurvey2003}. 
We write the probability that the new node $j$ links to an earlier node $i$ as
\begin{equation} \label{eq:model}
\Pi_{j \rightarrow i} = c_j^{-1} \frac{k_i |\r_j - \r_i| ^{-\alpha}}{j}
\end{equation}
where $k_i$ is the degree of the node $i$, $|\r_j - \r_i|$ is the Euclidean distance between $i$ and $j$, and $c_j$ is the normalization constant
\begin{equation}\label{eq:Ti}
c_j \equiv \frac{1}{j} \sum_{i'<j} \frac{k_{i'}}{m} |\r_j - \r_{i'}|^{-\alpha} 
\end{equation}

Equation (\ref{eq:model}--\ref{eq:Ti}) captures the coexistence of two effects that are separately known to influence the evolution of spatial networks: (a) the $k$-dependence captures preferential attachment, whose presence has been explicitly documented in numerous spatial networks \cite{yook2002modeling, manna2002modulated, xulvi2002evolving}; (b) the $|\r_j - \r_i| ^{-\alpha}$ term captures the role of the physical distance in a node's decision where to link.  Note that in this model we are interested in networks in real spaces, and only consider the background space to be Euclidean with its dimensionality two or three. Therefore, the model explored here is distinct from network embedding approaches, where the background space is determined by link connectivity \cite{serrano2008self,krioukov2009curvature,krioukov2010hyperbolic}. Related work also includes the work of Rozenfeld et al. \cite{rozenfeld2002scale}, who explored the spatial embedding of scale-free networks. Given that the work takes the scale-freeness as a modelling input, it is not designed to address why and under what conditions can scale-free networks emergence in spatial systems. Also, Warren et al. \cite{warren2002geography} and Hermann et al. \cite{herrmann2003connectivity} predict the emergence of a scale-free network in a spatial system, achieving this by choosing carefully the density distribution of the spatial placement of the nodes. They successfully show how node density determines the degree distribution if nodes link to all nodes within a predefined radius R. Here we want to offer a more general analysis, not placing constrains on the node density distribution—but rather explore the role of an arbitrary spatial density of nodes on the degree distribution of a growing network.

The model  (\ref{eq:model}--\ref{eq:Ti}) was introduced over a decade ago to capture the evolution of the Internet \cite{yook2002modeling}, and it has been subject to extensive numerical analysis for $D=1$, $2$ \cite{manna2002modulated, xulvi2002evolving}. Indeed, for $\alpha = 0$ the model reduces to the scale-free model \cite{barabasi1999emergence,BRdiam}, and nodes connect to each other without geographical restrictions. In this case, the degree distribution follows $P(k) \sim k^{-\gamma}$ with degree exponent $\gamma = 3$. When $0\le \alpha<2D$, which is the case for all known spatial networks, the distribution of the pairwise distances is well approximated by (see Fig. \ref{fig3}a)
\begin{equation} \label{Ee:Pr}
P(r) \sim r^{-\alpha} r^{D-1} = r^{-\delta},
\end{equation}
where the {\it distance exponent} is $\delta = 1 + \alpha - D$. If the spatial density of the nodes follows a fractal pattern, then $D$ is replaced by the fractal dimension $D_f$. For $\alpha > 2D_f$, we find $\delta = 1+D_f$, i.e. $\delta$ is independent of $\alpha$, as now the geographic restriction dominates, forcing each node to connect only to its closest neighbors (see SI Section 2D).

\begin{figure}[!htb]
\centering
\resizebox{\figsizeB}{!}
{\includegraphics{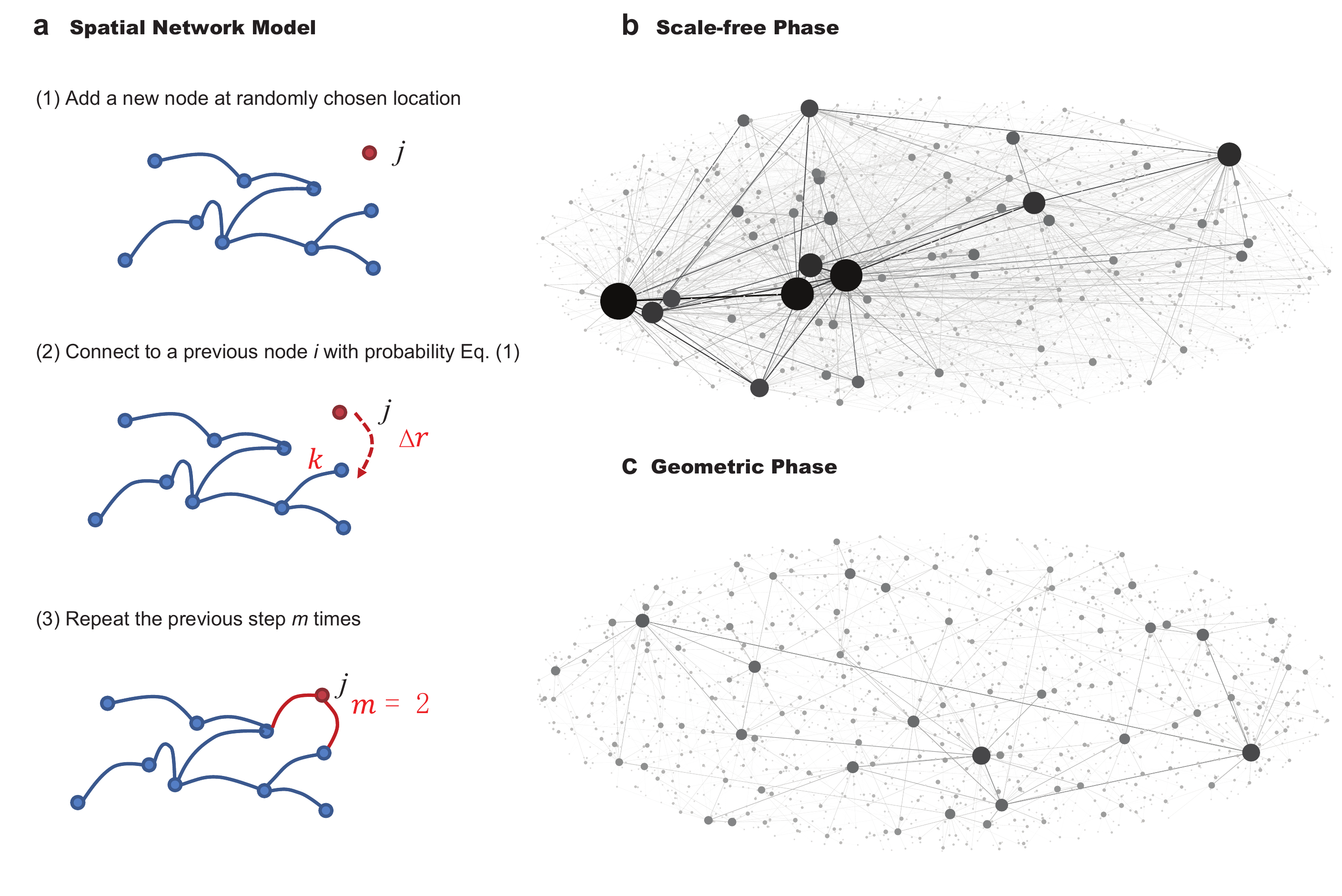}}
\caption{\textbf{Spatial network model}.  \textbf{(a)} We start from a set of nodes embedded in a $D$-dimensional space (shown here for $D=2$). In step (1) we add a new node $j$ at the randomly chosen position $\r$,  whose location is determined by the population density $\rho(\r)$; (2) We choose an existing node $i$ randomly with probability proportional to $k_i r_{ij}^{-\alpha}$, where $k_i$ and $r_{ij}$ represent node $i$'s degree and its distance from $j$; (3) We repeat step (2) $m$ times, generating $m$ new links between the node $j$ and preexisting nodes. \textbf{(b-c)} A network with $N = 2,000$ nodes generated by the model described in (a) for $m = 2$ on a two-dimensional plane ($D = D_f = 2$) where each node size is proportional to its degree. The network shown in (b) has $\alpha = 1< D_f$ and so is in the scale-free phase; its topology is dominated by a few large hubs. For (c) we have $\alpha = 3 > D_f$, so the resulting network is in the geometric phase. }
\label{fig2}
\end{figure}

Despite repeated attempts to characterize, using simulation, the model, its fundamental properties have remained elusive: How does the network topology change as we modify $\alpha$? Is the transition from a scale-free ($\alpha = 0$) to a planar network ($\alpha\rightarrow\infty$) a gradual process, or an abrupt one at some critical $\alpha_c$? How does this transition affect the various network characteristics? What is the form of the resulting $P(k)$?  Numerical simulations, due to size limitations, have failed to shed light on these questions \cite{manna2002modulated, xulvi2002evolving}, yet they offered important insights, that guide our work.
In one dimension R. Xulvi-Brunet et al. conjectured a phase-transition at $\alpha_c = 1$ based on numerical simulations \cite{xulvi2002evolving}, but the behavior for $D=2,3$,  the cases of direct practical relevance, could not be clarified numerically. The lack of an analytical framework to accurately predict the properties of this minimal model and to guide our expectations lies at the heart of all of these numerical limitations. The absence of such a framework also limits the fundamental understanding of spatial networks.
Here we offer an \emph{exact mathematical proof} that the model (\ref{eq:model}--\ref{eq:Ti}) undergoes a phase transition at $\alpha_c = D_f$ \cite{goltsev2003critical,dorogovtsev2008critical}. Our main advance is the demonstration that different continuum theories describe the evolution of the network on the two sides of $\alpha_c$, that allows us to predict the rich topological characteristics of spatial networks (see Fig. \ref{fig2}). A key finding is that the degree distribution and the degree exponent depend on the spatial node density $\rho(\r)$.  

To unveil the origin of the phase transition characterizing spatial networks, we must inspect the normalization constant (\ref{eq:Ti})
\begin{equation}\label{eq:int}
c_j \approx \int_{\rminj}^{\rmax} r^{-\alpha} r^{D_f-1} dr = \frac{\rmax^{D_f-\alpha}-\rminj^{D_f-\alpha}}{D_f-\alpha},
\end{equation}
where $\rminj$ and $\rmax$ are the distances to the closest and the most distant nodes from node $j$, respectively, and $r^{D_f-1} dr$ is (up to constants) the density of nodes at distance $r$ from the point $\r_j$. Note that the higher the node density, the smaller $\rminj$. 

(i) For $\alpha < D_f$ (\ref{eq:int}) converges in the limit $\rminj \rightarrow 0$, predicting $c_j \approx \rmax^{D_f-\alpha}$, i.e. $c_j$ is a constant independent of the choice of node $j$ (SI Section 2C). In this case, the spatial constraints have only limited impact and (\ref{eq:model}) is dominated by preferential attachment. Hence the network's connection pattern is global, prompting us to call this regime the \emph{scale-free phase}. 

(ii) In contrast, for $\alpha > D_f$, the integral (\ref{eq:int}) diverges as $c_j \approx \rminj^{D_f-\alpha}$ in the $\rminj \rightarrow 0$ limit. Since $\rminj$ depends on the local node density around node $j$, in this regime  the physical distance dominates the system's behavior and (\ref{eq:model}) captures a phase dominated by local density fluctuations, prompting us to call it the \emph{geometric phase}. 
Note that in (\ref{eq:int}) and hereafter $D_f$ denotes the fractal dimension of the node locations in the space, and not the fractal dimension of the resulting network \cite{song2005self}.

While the above heuristic argument based on the convergence of (\ref{eq:Ti}) ignores the role of the stochastic fluctuations in the degree evolution, in the SI Section 2E we \emph{prove rigorously that the phase transition at $\alpha = D_f$ is in fact exact}. To be specific, we demonstrate that several key network measures,  including the degree distribution, behave a distinctly different behavior, on the two sides of a sharp boundary in the parameter space, including (in the $n\rightarrow \infty$ limit) arbitrarily close to that boundary. The existence of this phase transition, with the associated proof, represents our first key result. Yet, the true impact of this transition can be appreciated only after predicting the fundamental network characteristics in the two phases separated by the phase transition, as we discuss next.

\emph{Degree Distribution.}  The degree distribution is known to affect most network properties, from network robustness to spreading patterns, prompting us to first inspect the impact of the phase transition on $P(k)$. In the \emph{scale-free phase} ($\alpha < D_f$), $c_j$ is finite and the degree of node $i$ evolves as
\begin{equation} \label{Eq:dkdN}
\frac{d k_i}{d t} = \sigma(\r_i)\frac{k_i}{t},
\end{equation}
where
\begin{equation}\label{Eq:norm}
\sigma(\r_i) \equiv m \int \frac{|\r - \r_i|^{-\alpha}}{ \sum_{i'\le t} k_{i'} |\r-\r_{i'}|^{-\alpha}/t} \rho(\r) d\r
\end{equation}
is the normalization factor, and the time step $t$ is equivalent to the current network size $j$ in (\ref{eq:model}).  Note that in (\ref{Eq:norm}) the numerator $\sum_{i'\le t} k_{i'} |r-r_{i'}|^{-\alpha}/t$ does depend on the time $t$.  However, this term equals to the weighted average degree $< k_{i} |r-r_{i}|^{-\alpha} >$ up to time $t$, that converges rapidly to a constant as $t$ increases. A more detailed discussion of this convergence is offered in SI 2C (Eq. S9-S10), where we track the time-dependency of the numerator, and in S2.3.3.3 where we finally find the stationary solution of the numerator for $t\rightarrow \infty$. As we show in SI 2C, (\ref{Eq:dkdN}) predicts
\begin{equation} \label{Eq:dkdt}
k_i \sim (t/i)^{\sigma(\r_i)},
\end{equation}
and a power law degree distribution $P(k) \sim k^{-\gamma}$, where $\gamma = 1 + 1/\sup_{\r} \sigma(\r)$. Here $\sup_{\r} \sigma(\r)$ is the maximum of $\sigma$ over all positions $\r$ in the background space. Equations (\ref{Eq:dkdN}-\ref{Eq:dkdt}) represent our second key result. If the nodes are uniformly distributed in a uniform, unbounded space ($\rho(\r) = const$), we have $\sigma = 1/2$ (see SI Section 2C), $k_i \sim (t/i)^{1/2}$ and $\gamma =3$, recovering the known degree exponent of the scale-free model. Equation (\ref{Eq:dkdt}) predicts, however, that a non-uniform population density, $\rho(\r)$, can alter the degree exponent $\gamma$. Remarkably, this could also result from straightforward boundary effects. Consider, for example, a uniform density $\rho(\r)$ on a bounded region, such as a square. The likelihood of connecting to nodes near the edge is slightly less than to nodes close to the center, leading to a non-uniform $\sigma(\r)$. Previously, such finite size effects, that gradually disappear as the number of nodes grows, were not expected to affect the asymptotic scaling exponents \cite{BRsurvey2003}. Our theory predicts, however, that in spatial networks the non-uniform $\sigma(\r)$ changes the degree exponent to $\gamma = 2.803$ for $\alpha = 1$ (SI Section 2C).
Figure \ref{fig3}d plots the analytical prediction of the average degree in different spatial regimes in this case, revealing the inhomogeneity generated by the boundary effect (SI section 2C). To obtain numerical evidence for this unexpected prediction, in Figure \ref{fig3}c we compare $\gamma$ for $2^{30} \approx 1$B nodes uniformly distributed in a square with and without periodic boundary conditions, where the former corresponds to a uniform $\sigma(\r)$, hence we expect $\gamma=3$, and the latter has a non-uniform $\sigma(\r)$, expecting a smaller $\gamma$. The simulation, spanning four decades in degrees, fully confirms the altered exponent $\gamma$ predicted analytically.

This effect is also important because most processes that influence the degree exponent, from link deletion to rewiring, are known to increase $\gamma$, rather than decrease it   \cite{barabasi2016network,ghoshal2013uncovering}. In contrast spatial inhomogeneity decreases $\gamma$ below the critical value $\gamma =3$, leading to a diverging second moment, which in turn induces anomalous robustness \cite{cohen2000resilience} and a vanishing epidemic threshold \cite{pastor2001epidemic}. Our calculation suggests that these desirable (robustness) or undesirable (vanishing epidemic threshold) features of scale-free networks can be induced by spatial inhomogeneities.

Turning now to the \emph{geometric phase}, we find that for $\alpha > D_f$ the integral (\ref{eq:Ti}) diverges at small length scales, forcing us to systematically consider the impact of the small distance cutoff $\rminj$. Note that for a new node $j$ the rescaled normalization factor $jc_j$ is within a constant factor of $k_ir_{ij}^{-\alpha}$, where $i$ is the closest node to $j$. Since $r_{ij}$ is of order $j^{-1/D_f}$ (the more nodes we have, the closer will be a new node to an preexisting node), we have $jc_j \approx k_ij^{\alpha/D_f}$. Therefore, for node $j$ to link to node $i$, the distance $r_{ij}$ must fall within
\begin{equation}
 r_{ij} = k_i^{1/\alpha} j^{-1/D_f},
\end{equation}
leading to the upper bound (up to a constant factor) on the probability of connecting to node $i$ of (see SI section 2C)
\begin{equation}
\Pi_{i} = O(r_{ij}^{D_f}) = O(k_i^{D_f/\alpha}/j).
\end{equation}

Consequently, the growth of a node's degree in the geometric phase is not governed by (\ref{Eq:dkdN}), but by sublinear preferential attachment \cite{krapivsky2000connectivity}:
\begin{equation}
\frac{d k_i}{d t} \propto \frac{k_i^{D_f/\alpha}}{t},
\end{equation}
predicting that in the geometric phase $P(k)$ follows a stretched exponential, 
\begin{equation} \label{eq:PKG}
P_>(k) \sim \exp(- A k^{1-D_f/\alpha}),
\end{equation}
where $A$ is a normalization constant. This is our third key result, predicting that spatial separation overcomes the effect of preferential attachment, inducing a stretched exponential cutoff in the degree distribution.

\begin{figure}[!htb]
\centering
\resizebox{\figsizeA}{!}
{\includegraphics{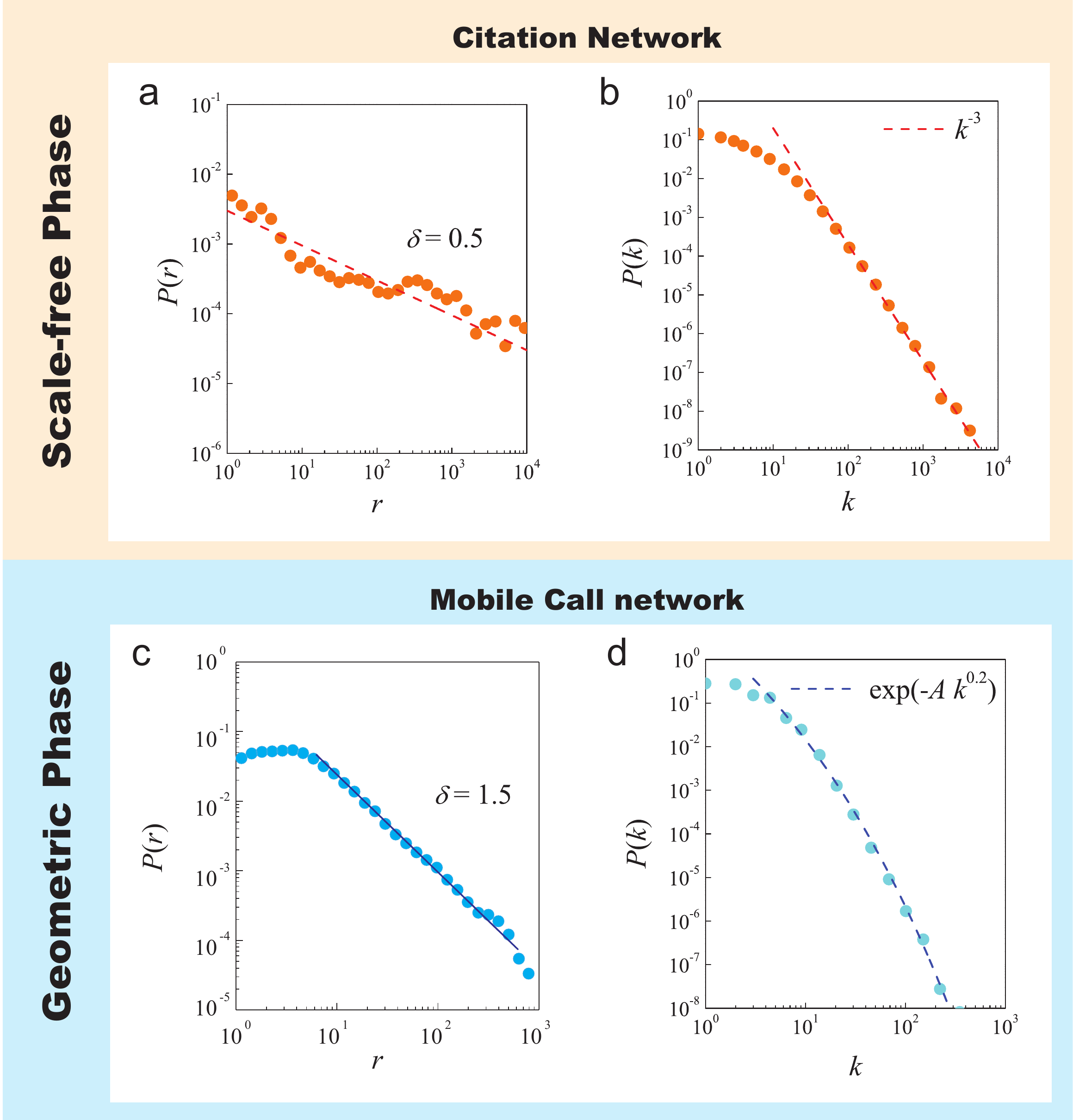}}
\caption{\textbf{Empirical results}. The degree distribution $P(k)$ and spatial correlation function
$P(r)$ for \textbf{(a-b)} the citation network and \textbf{(c-d)} the mobile phone network. The distance distribution follows $P(r) \sim r^{-\delta}$ for both networks, with $\delta = 0.5$ for citations and $\delta = 1.5$ for mobile phones. Hence we predict that the citation network belongs to the scale-free phase ($\alpha<D_f$) and the call network to the geometric phase ($\alpha > D_f$). Consequently, the resulting degree distribution should follow a power law  for the citation network (b) and a stretched exponential for the mobile phone network (d). The dashed lines show the analytic predictions for each quantity.}
\label{fig1}
\end{figure}

\begin{figure}[!htb]
\centering
\resizebox{\figsizeC}{!}
{\includegraphics{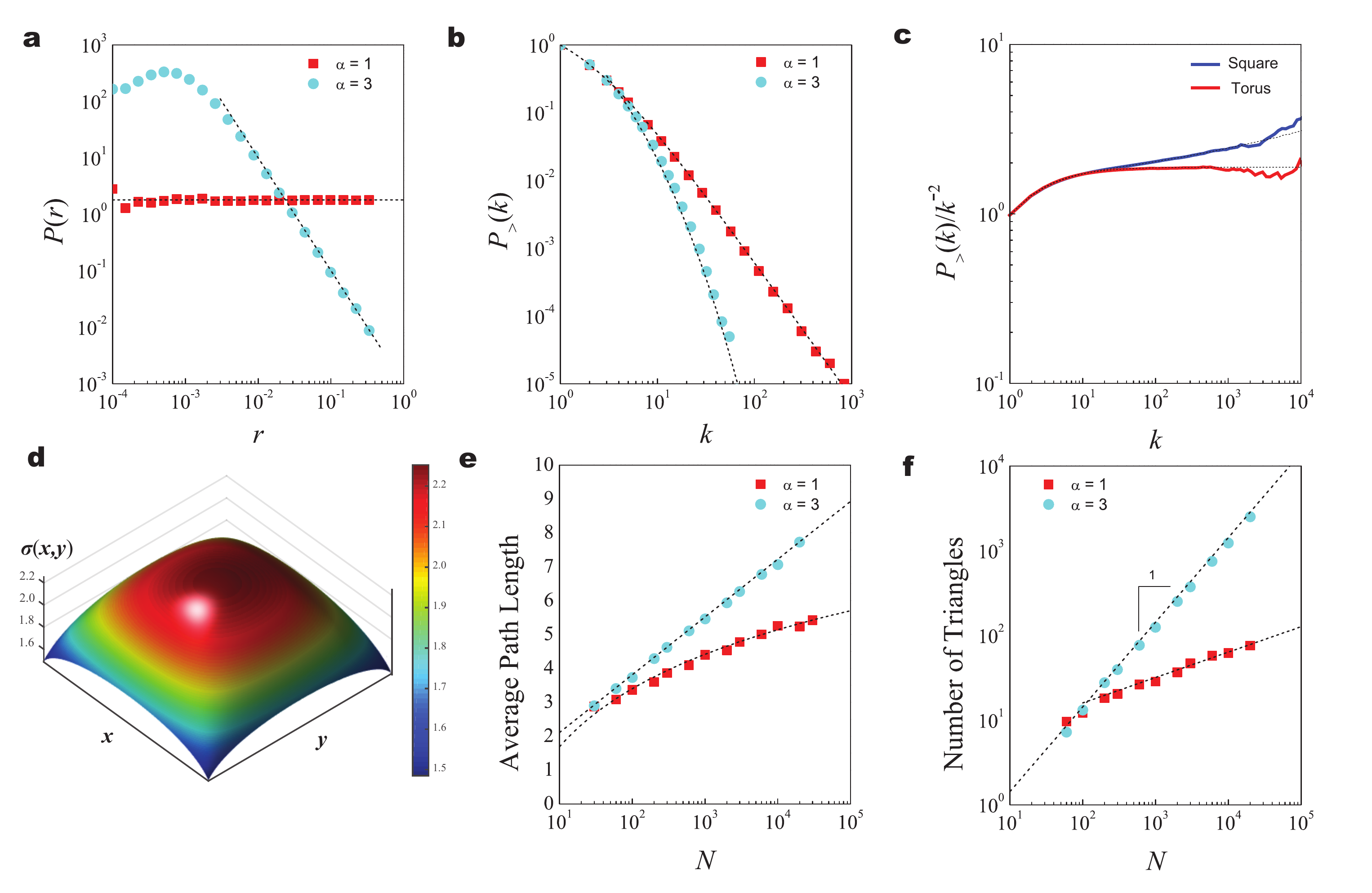}}
\caption{\textbf{Testing the theoretical predictions. } 
\textbf{(a)} Distance distribution for different $\alpha$ for uniform $\rho(\r)$ with $D = 2$ (hence $D_f = 2$), $N = 30,000$ and $m = 2$. The plot confirms the theoretical prediction (\ref{Ee:Pr}) that $P(r) \sim r^{-\delta}$ with $\delta = 1 + \alpha - D_f$, shown as dashed lines. Indeed, for $\alpha =1$ we expect $\delta = 0$, observing that indeed $P(r)$ does not depend on $r$, whereas for $\alpha = 3$ $P(r)$ decays as $1/r^2$ as predicted.
\textbf{(b)} Degree distribution for a uniform $\rho(\r)$ in $D=2$, for $\alpha = 1$ and $3$. For $\alpha = 1$  $P(k)$ is expected to follow a pure power law, in line with the numerical results (scale-free regime). For $\alpha = 3$ we are in the geometric phase, hence $P(k)$ follows a stretched exponential. The dashed lines correspond to the theoretical predictions $P(k) \sim k^{-3}$ and $\exp(-Ak^{1-D_f/\alpha})$ for $\alpha = 1$ and $3$, respectively. 
\textbf{(c)} Cumulative degree distribution $P_>(k)$ normalized by $k^{-2}$ for networks generated with $\alpha=1$ using a uniform $\rho(\r)$ on a $D=2$ plane with $N = 2^{30}$. The square has a fixed boundary, while the torus wraps around, making a homogeneous space in which all points are equivalent.
The theory predicts $P_>(k) \sim k^{-(\gamma-1)}$ for the scale-free phase, hence $P_>(k)/k^{-2} \sim k^{3-\gamma}$. Consequently the log-log plot captures the deviation of the degree exponent from $\gamma=3$. The horizontal line for the torus confirms that $\gamma = 3$ in this homogeneous space. Introducing a boundary (square) gives $\gamma < 3$. The dashed lines represent the theoretical predictions (see SI Section 2C), which apart from fluctuations, are indistinguishable from the simulation results. 
\textbf{(d)} The theoretical prediction for the degree density $\sigma(\r)$ for a uniform $\rho(\r)$ and $\alpha = 1$ on a $D=2$ ($D_f = 2$) plane with fixed square boundary (SI Section 2D), indicating that the boundary effect generates attractiveness inhomogeneity for different spatial regimes, hence $\gamma$ is different from $3$.  
\textbf{(e)} Average path length vs. network size for networks generated with uniform $\rho(\r)$ in $D=2$ with boundary ($m=2$ and $\gamma < 3$), finding a logarithmic dependence for $\alpha = 3$ and a double logarithmic dependence for $\alpha = 1$, in line with the prediction (\ref{eq:l}). 
\textbf{(f)} The number of triangles vs. network size for networks with a uniform $\rho(\r)$ in $D=2$ ($D_f = 2$) with different $\alpha$, finding that the number of triangles scales linearly for $\alpha > D_f = 2$ and sublinearly for $\alpha < D_f = 2$, in agreement with the prediction (\ref{eq:C}).  }
\label{fig3}
\end{figure}

To demonstrate that these predictions have direct implications for real systems, we focus on two networks for which we can experimentally resolve both the underlying network structure and the geographical layout (SI Section 1). The first system is the citation network of 463,348 papers published between 1893 and 2010, extracted from the Physical Review corpus that spans all areas of physics \cite{redner2005citation}. Here a node is a scientific publication and a link captures a citation from paper $i$ to paper $j$; the distance $r_{ij}$ captures the geographical separation of the first author affiliations. Figure~\ref{fig1}a shows the observed $P(r)$ for the citation network, finding that it follows Eq. (\ref{Ee:Pr}) with $\delta \approx 0.5$ for four orders of magnitude. Relation (\ref{Ee:Pr}) predicts $\delta = 1 + \alpha - D_f$, implying that for the citation network $\alpha < D_f$, hence we are in the scale-free phase and $P(k)$ should be fat tailed. This prediction is fully confirmed by the measurements: we find in line with many previous results \cite{redner1998popular, redner2005citation} that $P(k)$ has a power law tail with $\gamma = 3$ (Fig.~\ref{fig1}b).

The second system is the mobile phone network that captures the call patterns of about 3 million anonymized European mobile phone users during a one-year period. Here nodes represent mobile phones and each link corresponds to a direct call between two users; $r_{ij}$ captures the geographical distance between users' $i$ and $j$ most likely locations. The measurement of $P(r)$ indicates $\delta \approx 1.5$ (Fig.~\ref{fig1}c), hence according to Eq. (\ref{Ee:Pr}) we have $\alpha > D_f$, predicting that the mobile phone network belongs to the geometric phase. We therefore predict a stretched exponential (\ref{eq:PKG}) for $P(k)$, which not only offers an excellent fit to the data (Fig. \ref{fig1}d), but also resolves a long-standing mystery of mobile phone networks: earlier power law fits of $P(k)$ reported an exceptionally large degree exponent $\gamma \approx 8.4$ \cite{onnela2007structure}. Yet, as in this system spatial effects play an important role, the proper $P(k)$ is expected to follow $P_>(k) \sim \exp(-A k^\beta)$ with $\beta \approx 0.2\pm 0.1$ (Fig. \ref{fig1}d). Equation (\ref{eq:PKG}) also predicts a scaling identity $\beta = 1-D_f/\alpha$. To test this relationship, we measured the fractal dimension $D_f$ of the mobile phone users from $\rho(\r)$, finding $D_f \approx 1.0$. Together with Eq. (\ref{Ee:Pr}), we estimate $\alpha = \delta + D_f - 1 \approx 1.5$, which in turn predicts $\beta \approx 0.3\pm 0.05$, consistent within the error bars with the empirically observed $\beta \approx 0.2\pm 0.1$.

\emph{Average Path Length.} The predicted phase transition alters the path length structure of the network as well. Indeed, we analytically derived the average path length $\left<l\right>$ for the model (SI Section 2D), obtaining
\begin{equation}\label{eq:l}
\left<l \right> \sim \left\{
\begin{array}{ll}
\ln N, &m = 1\mathrm{\ or\ }\alpha > D_f\\
\ln \ln N, &m > 1\mathrm{\ and \ }\alpha < D_f\mathrm{\ and \ }\gamma < 3\\
\frac{\ln N}{\ln \ln N}, &m > 1\mathrm{\ and \ }\alpha < D_f\mathrm{\ and \ }\gamma = 3\\
\end{array}
\right. .
\end{equation}
Equation (\ref{eq:l}) makes several remarkable predictions. First, in the geometric phase, where nodes connect predominantly to their nearest neighbours, the average path length is expected to scale polynomially. Yet, we predict that the geometric phase displays the small world property for all $m$ values. For the scale-free phase, as long as when the underlying network is not a tree ($m>1$), we find either an ultra-small world property ($\ln\ln N$) or a double logarithmic correction for $\gamma =3$ \cite{BRdiam}. Prediction (\ref{eq:l}) is again confirmed by numerical simulations (Fig. \ref{fig3}e). 

\emph{Clustering.} The predicted phase transition affects local clustering as well. For $\alpha > D_f$ a positive fraction of links connect nodes to their closest or second closest neighbors. Therefore we expect a finite fraction of  triangles, meaning that the number $T$ of triangles is of order $N$. In contrast for $\alpha < D_f$ the probability $\Pi_{ijk}$ to form a triangle among nodes $i<j<k$ follows
\begin{equation} \label{Eq: CL}
\Pi_{ijk} = \Pi_{ij}\Pi_{ik}\Pi_{jk},
\end{equation}
where \begin{equation}
\Pi_{ij}\equiv \frac{(j/i)^{\sigma(\r_i)}r_{ij}^{-\alpha}}{\sum_{s<j}(j/s)^{\sigma(\r_s)}r_{sj}^{-\alpha}}
\end{equation}
is the probability connecting nodes $i$ and $j$. By averaging (\ref{Eq: CL}) over all nodes, we find
\begin{equation}\label{eq:C}
T \approx \left\{
\begin{array}{ll}
N, &\alpha > D_f\\
N^{\theta(\alpha, D_f)+o(1)}, &\alpha < D_f
\end{array}
\right. ,
\end{equation}
where $\theta(\alpha, D_f) < 1$, its precise value depending both on $\alpha$ and the dimension $D_f$ of the space (see SI Section 2D). Equation (\ref{eq:C}) predicts that in the geometric phase the networks show a high degree of local clustering. In the scale-free phase, however, the networks have fewer triangles. The prediction (\ref{eq:C}) is again confirmed by the numerical simulation (Fig. \ref{fig3}f). Furthermore, we find that despite the fact that the citation network is much denser than the mobile phone network ($\left<k\right> \approx 20.5$ for the citation network and $5.5$ for the mobile phone network), the ratio between the number of triangles and network size $T/N$ for both networks is comparable, i.e. $T/N \approx  0.240$ for citation ($0.00107258$ for random) and $0.215$ for mobile phone ($0.000540129$ for random). Indeed, by comparing to randomized networks with the same degree distribution, we find that the relative triangle frequency $T/T_{rand}$ in the mobile phone network (geometric phase) is twice that observed in the citation network (scale-free phase),  indicating that the prediction (\ref{eq:C}), that the geometric phase shows more local clustering than the scale-free phase, agrees qualitatively with the empirical observations.

In summary, our results fill in an important gap for the emergence and the topology of spatial networks. Firstly, our approach offers a rigorous link between the background geometry and the network topology, documenting a phase transition that separates two phases with distinct topological characteristics, governed by different equations. Note however, that despite the highly predictive results, the presented model is, like almost all non-trivial models in statistical physics, still far from being completely solved, leaving us number of unanswered questions for future investigation. For example, what are critical exponents of the observed phase transition? Nevertheless, the observed phases, whose direct relevance to real systems is demonstrated through empirical data, show remarkable differences in network characteristics such as the degree distribution, the clustering coefficient and the path lengths.  While more sophisticated spatial network models have to incorporate many additional system-dependent details, we believe that the universality classes uncovered here will retain their direct relevance, capable of guiding the investigation of real systems. As our understanding of spatial network deepens with the emergence of new and increasingly detailed data on both the geographical and the network sides, such universal phases offer a springboard towards a deeper mechanistic understanding of complex networks embedded in a space. As such our results are of direct relevance from transportation to understanding human connections.

\newpage
\bibliography{spatial}

%merlin.mbs apsrev4-1.bst 2010-07-25 4.21a (PWD, AO, DPC) hacked
%Control: key (0)
%Control: author (8) initials jnrlst
%Control: editor formatted (1) identically to author
%Control: production of article title (-1) disabled
%Control: page (0) single
%Control: year (1) truncated
%Control: production of eprint (0) enabled
\begin{thebibliography}{40}%
\makeatletter
\providecommand \@ifxundefined [1]{%
 \@ifx{#1\undefined}
}%
\providecommand \@ifnum [1]{%
 \ifnum #1\expandafter \@firstoftwo
 \else \expandafter \@secondoftwo
 \fi
}%
\providecommand \@ifx [1]{%
 \ifx #1\expandafter \@firstoftwo
 \else \expandafter \@secondoftwo
 \fi
}%
\providecommand \natexlab [1]{#1}%
\providecommand \enquote  [1]{``#1''}%
\providecommand \bibnamefont  [1]{#1}%
\providecommand \bibfnamefont [1]{#1}%
\providecommand \citenamefont [1]{#1}%
\providecommand \href@noop [0]{\@secondoftwo}%
\providecommand \href [0]{\begingroup \@sanitize@url \@href}%
\providecommand \@href[1]{\@@startlink{#1}\@@href}%
\providecommand \@@href[1]{\endgroup#1\@@endlink}%
\providecommand \@sanitize@url [0]{\catcode `\\12\catcode `\$12\catcode
  `\&12\catcode `\#12\catcode `\^12\catcode `\_12\catcode `\%12\relax}%
\providecommand \@@startlink[1]{}%
\providecommand \@@endlink[0]{}%
\providecommand \url  [0]{\begingroup\@sanitize@url \@url }%
\providecommand \@url [1]{\endgroup\@href {#1}{\urlprefix }}%
\providecommand \urlprefix  [0]{URL }%
\providecommand \Eprint [0]{\href }%
\providecommand \doibase [0]{http://dx.doi.org/}%
\providecommand \selectlanguage [0]{\@gobble}%
\providecommand \bibinfo  [0]{\@secondoftwo}%
\providecommand \bibfield  [0]{\@secondoftwo}%
\providecommand \translation [1]{[#1]}%
\providecommand \BibitemOpen [0]{}%
\providecommand \bibitemStop [0]{}%
\providecommand \bibitemNoStop [0]{.\EOS\space}%
\providecommand \EOS [0]{\spacefactor3000\relax}%
\providecommand \BibitemShut  [1]{\csname bibitem#1\endcsname}%
\let\auto@bib@innerbib\@empty
%</preamble>
\bibitem [{\citenamefont {Barth{\'e}lemy}(2011)}]{barthelemy2011spatial}%
  \BibitemOpen
  \bibfield  {author} {\bibinfo {author} {\bibfnamefont {M.}~\bibnamefont
  {Barth{\'e}lemy}},\ }\href@noop {} {\bibfield  {journal} {\bibinfo  {journal}
  {Physics Reports}\ }\textbf {\bibinfo {volume} {499}},\ \bibinfo {pages} {1}
  (\bibinfo {year} {2011})}\BibitemShut {NoStop}%
\bibitem [{\citenamefont {Liben-Nowell}\ \emph {et~al.}(2005)\citenamefont
  {Liben-Nowell}, \citenamefont {Novak}, \citenamefont {Kumar}, \citenamefont
  {Raghavan},\ and\ \citenamefont {Tomkins}}]{liben2005geographic}%
  \BibitemOpen
  \bibfield  {author} {\bibinfo {author} {\bibfnamefont {D.}~\bibnamefont
  {Liben-Nowell}}, \bibinfo {author} {\bibfnamefont {J.}~\bibnamefont {Novak}},
  \bibinfo {author} {\bibfnamefont {R.}~\bibnamefont {Kumar}}, \bibinfo
  {author} {\bibfnamefont {P.}~\bibnamefont {Raghavan}}, \ and\ \bibinfo
  {author} {\bibfnamefont {A.}~\bibnamefont {Tomkins}},\ }\href@noop {}
  {\bibfield  {journal} {\bibinfo  {journal} {Proceedings of the National
  Academy of Sciences}\ }\textbf {\bibinfo {volume} {102}},\ \bibinfo {pages}
  {11623} (\bibinfo {year} {2005})}\BibitemShut {NoStop}%
\bibitem [{\citenamefont {Lambiotte}\ \emph {et~al.}(2008)\citenamefont
  {Lambiotte}, \citenamefont {Blondel}, \citenamefont {de~Kerchove},
  \citenamefont {Huens}, \citenamefont {Prieur}, \citenamefont {Smoreda},\ and\
  \citenamefont {Van~Dooren}}]{lambiotte2008geographical}%
  \BibitemOpen
  \bibfield  {author} {\bibinfo {author} {\bibfnamefont {R.}~\bibnamefont
  {Lambiotte}}, \bibinfo {author} {\bibfnamefont {V.~D.}\ \bibnamefont
  {Blondel}}, \bibinfo {author} {\bibfnamefont {C.}~\bibnamefont
  {de~Kerchove}}, \bibinfo {author} {\bibfnamefont {E.}~\bibnamefont {Huens}},
  \bibinfo {author} {\bibfnamefont {C.}~\bibnamefont {Prieur}}, \bibinfo
  {author} {\bibfnamefont {Z.}~\bibnamefont {Smoreda}}, \ and\ \bibinfo
  {author} {\bibfnamefont {P.}~\bibnamefont {Van~Dooren}},\ }\href@noop {}
  {\bibfield  {journal} {\bibinfo  {journal} {Physica A: Statistical Mechanics
  and its Applications}\ }\textbf {\bibinfo {volume} {387}},\ \bibinfo {pages}
  {5317} (\bibinfo {year} {2008})}\BibitemShut {NoStop}%
\bibitem [{\citenamefont {Pan}\ \emph {et~al.}(2012)\citenamefont {Pan},
  \citenamefont {Kaski},\ and\ \citenamefont {Fortunato}}]{pan2012world}%
  \BibitemOpen
  \bibfield  {author} {\bibinfo {author} {\bibfnamefont {R.~K.}\ \bibnamefont
  {Pan}}, \bibinfo {author} {\bibfnamefont {K.}~\bibnamefont {Kaski}}, \ and\
  \bibinfo {author} {\bibfnamefont {S.}~\bibnamefont {Fortunato}},\ }\href@noop
  {} {\bibfield  {journal} {\bibinfo  {journal} {Scientific reports}\ }\textbf
  {\bibinfo {volume} {2}} (\bibinfo {year} {2012})}\BibitemShut {NoStop}%
\bibitem [{\citenamefont {Adamic}\ and\ \citenamefont
  {Adar}(2005)}]{adamic2005search}%
  \BibitemOpen
  \bibfield  {author} {\bibinfo {author} {\bibfnamefont {L.}~\bibnamefont
  {Adamic}}\ and\ \bibinfo {author} {\bibfnamefont {E.}~\bibnamefont {Adar}},\
  }\href@noop {} {\bibfield  {journal} {\bibinfo  {journal} {Social Networks}\
  }\textbf {\bibinfo {volume} {27}},\ \bibinfo {pages} {187} (\bibinfo {year}
  {2005})}\BibitemShut {NoStop}%
\bibitem [{\citenamefont {Yook}\ \emph {et~al.}(2002)\citenamefont {Yook},
  \citenamefont {Jeong},\ and\ \citenamefont
  {Barab{\'a}si}}]{yook2002modeling}%
  \BibitemOpen
  \bibfield  {author} {\bibinfo {author} {\bibfnamefont {S.-H.}\ \bibnamefont
  {Yook}}, \bibinfo {author} {\bibfnamefont {H.}~\bibnamefont {Jeong}}, \ and\
  \bibinfo {author} {\bibfnamefont {A.-L.}\ \bibnamefont {Barab{\'a}si}},\
  }\href@noop {} {\bibfield  {journal} {\bibinfo  {journal} {Proceedings of the
  National Academy of Sciences}\ }\textbf {\bibinfo {volume} {99}},\ \bibinfo
  {pages} {13382} (\bibinfo {year} {2002})}\BibitemShut {NoStop}%
\bibitem [{\citenamefont {Masucci}\ \emph {et~al.}(2009)\citenamefont
  {Masucci}, \citenamefont {Smith}, \citenamefont {Crooks},\ and\ \citenamefont
  {Batty}}]{masucci2009random}%
  \BibitemOpen
  \bibfield  {author} {\bibinfo {author} {\bibfnamefont {A.}~\bibnamefont
  {Masucci}}, \bibinfo {author} {\bibfnamefont {D.}~\bibnamefont {Smith}},
  \bibinfo {author} {\bibfnamefont {A.}~\bibnamefont {Crooks}}, \ and\ \bibinfo
  {author} {\bibfnamefont {M.}~\bibnamefont {Batty}},\ }\href@noop {}
  {\bibfield  {journal} {\bibinfo  {journal} {The European Physical Journal B}\
  }\textbf {\bibinfo {volume} {71}},\ \bibinfo {pages} {259} (\bibinfo {year}
  {2009})}\BibitemShut {NoStop}%
\bibitem [{\citenamefont {Bianconi}\ \emph {et~al.}(2009)\citenamefont
  {Bianconi}, \citenamefont {Pin},\ and\ \citenamefont
  {Marsili}}]{bianconi2009assessing}%
  \BibitemOpen
  \bibfield  {author} {\bibinfo {author} {\bibfnamefont {G.}~\bibnamefont
  {Bianconi}}, \bibinfo {author} {\bibfnamefont {P.}~\bibnamefont {Pin}}, \
  and\ \bibinfo {author} {\bibfnamefont {M.}~\bibnamefont {Marsili}},\
  }\href@noop {} {\bibfield  {journal} {\bibinfo  {journal} {Proceedings of the
  National Academy of Sciences}\ }\textbf {\bibinfo {volume} {106}},\ \bibinfo
  {pages} {11433} (\bibinfo {year} {2009})}\BibitemShut {NoStop}%
\bibitem [{\citenamefont {Srere}(2000)}]{srere2000macromolecular}%
  \BibitemOpen
  \bibfield  {author} {\bibinfo {author} {\bibfnamefont {P.~A.}\ \bibnamefont
  {Srere}},\ }\href@noop {} {\bibfield  {journal} {\bibinfo  {journal} {Trends
  in biochemical sciences}\ }\textbf {\bibinfo {volume} {25}},\ \bibinfo
  {pages} {150} (\bibinfo {year} {2000})}\BibitemShut {NoStop}%
\bibitem [{\citenamefont {Sayyid}\ and\ \citenamefont
  {Kalvala}(2016)}]{sayyid2016importance}%
  \BibitemOpen
  \bibfield  {author} {\bibinfo {author} {\bibfnamefont {F.}~\bibnamefont
  {Sayyid}}\ and\ \bibinfo {author} {\bibfnamefont {S.}~\bibnamefont
  {Kalvala}},\ }\href@noop {} {\bibfield  {journal} {\bibinfo  {journal}
  {Biosystems}\ }\textbf {\bibinfo {volume} {145}},\ \bibinfo {pages} {53}
  (\bibinfo {year} {2016})}\BibitemShut {NoStop}%
\bibitem [{\citenamefont {Bullmore}\ and\ \citenamefont
  {Sporns}(2009)}]{bullmore2009complex}%
  \BibitemOpen
  \bibfield  {author} {\bibinfo {author} {\bibfnamefont {E.}~\bibnamefont
  {Bullmore}}\ and\ \bibinfo {author} {\bibfnamefont {O.}~\bibnamefont
  {Sporns}},\ }\href@noop {} {\bibfield  {journal} {\bibinfo  {journal} {Nature
  Reviews Neuroscience}\ }\textbf {\bibinfo {volume} {10}},\ \bibinfo {pages}
  {186} (\bibinfo {year} {2009})}\BibitemShut {NoStop}%
\bibitem [{\citenamefont {Markov}\ \emph
  {et~al.}(2013{\natexlab{a}})\citenamefont {Markov}, \citenamefont
  {Ercsey-Ravasz}, \citenamefont {Van~Essen}, \citenamefont {Knoblauch},
  \citenamefont {Toroczkai},\ and\ \citenamefont
  {Kennedy}}]{markov2013cortical}%
  \BibitemOpen
  \bibfield  {author} {\bibinfo {author} {\bibfnamefont {N.~T.}\ \bibnamefont
  {Markov}}, \bibinfo {author} {\bibfnamefont {M.}~\bibnamefont
  {Ercsey-Ravasz}}, \bibinfo {author} {\bibfnamefont {D.~C.}\ \bibnamefont
  {Van~Essen}}, \bibinfo {author} {\bibfnamefont {K.}~\bibnamefont
  {Knoblauch}}, \bibinfo {author} {\bibfnamefont {Z.}~\bibnamefont
  {Toroczkai}}, \ and\ \bibinfo {author} {\bibfnamefont {H.}~\bibnamefont
  {Kennedy}},\ }\href@noop {} {\bibfield  {journal} {\bibinfo  {journal}
  {Science}\ }\textbf {\bibinfo {volume} {342}},\ \bibinfo {pages} {1238406}
  (\bibinfo {year} {2013}{\natexlab{a}})}\BibitemShut {NoStop}%
\bibitem [{\citenamefont {Markov}\ \emph
  {et~al.}(2013{\natexlab{b}})\citenamefont {Markov}, \citenamefont
  {Ercsey-Ravasz}, \citenamefont {Lamy}, \citenamefont {Gomes}, \citenamefont
  {Magrou}, \citenamefont {Misery}, \citenamefont {Giroud}, \citenamefont
  {Barone}, \citenamefont {Dehay}, \citenamefont {Toroczkai} \emph
  {et~al.}}]{markov2013role}%
  \BibitemOpen
  \bibfield  {author} {\bibinfo {author} {\bibfnamefont {N.~T.}\ \bibnamefont
  {Markov}}, \bibinfo {author} {\bibfnamefont {M.}~\bibnamefont
  {Ercsey-Ravasz}}, \bibinfo {author} {\bibfnamefont {C.}~\bibnamefont {Lamy}},
  \bibinfo {author} {\bibfnamefont {A.~R.~R.}\ \bibnamefont {Gomes}}, \bibinfo
  {author} {\bibfnamefont {L.}~\bibnamefont {Magrou}}, \bibinfo {author}
  {\bibfnamefont {P.}~\bibnamefont {Misery}}, \bibinfo {author} {\bibfnamefont
  {P.}~\bibnamefont {Giroud}}, \bibinfo {author} {\bibfnamefont
  {P.}~\bibnamefont {Barone}}, \bibinfo {author} {\bibfnamefont
  {C.}~\bibnamefont {Dehay}}, \bibinfo {author} {\bibfnamefont
  {Z.}~\bibnamefont {Toroczkai}},  \emph {et~al.},\ }\href@noop {} {\bibfield
  {journal} {\bibinfo  {journal} {Proceedings of the National Academy of
  Sciences}\ }\textbf {\bibinfo {volume} {110}},\ \bibinfo {pages} {5187}
  (\bibinfo {year} {2013}{\natexlab{b}})}\BibitemShut {NoStop}%
\bibitem [{\citenamefont {Ercsey-Ravasz}\ \emph {et~al.}(2013)\citenamefont
  {Ercsey-Ravasz}, \citenamefont {Markov}, \citenamefont {Lamy}, \citenamefont
  {Van~Essen}, \citenamefont {Knoblauch}, \citenamefont {Toroczkai},\ and\
  \citenamefont {Kennedy}}]{ercsey2013predictive}%
  \BibitemOpen
  \bibfield  {author} {\bibinfo {author} {\bibfnamefont {M.}~\bibnamefont
  {Ercsey-Ravasz}}, \bibinfo {author} {\bibfnamefont {N.~T.}\ \bibnamefont
  {Markov}}, \bibinfo {author} {\bibfnamefont {C.}~\bibnamefont {Lamy}},
  \bibinfo {author} {\bibfnamefont {D.~C.}\ \bibnamefont {Van~Essen}}, \bibinfo
  {author} {\bibfnamefont {K.}~\bibnamefont {Knoblauch}}, \bibinfo {author}
  {\bibfnamefont {Z.}~\bibnamefont {Toroczkai}}, \ and\ \bibinfo {author}
  {\bibfnamefont {H.}~\bibnamefont {Kennedy}},\ }\href@noop {} {\bibfield
  {journal} {\bibinfo  {journal} {Neuron}\ }\textbf {\bibinfo {volume} {80}},\
  \bibinfo {pages} {184} (\bibinfo {year} {2013})}\BibitemShut {NoStop}%
\bibitem [{\citenamefont {V{\'e}rtes}\ \emph {et~al.}(2012)\citenamefont
  {V{\'e}rtes}, \citenamefont {Alexander-Bloch}, \citenamefont {Gogtay},
  \citenamefont {Giedd}, \citenamefont {Rapoport},\ and\ \citenamefont
  {Bullmore}}]{vertes2012simple}%
  \BibitemOpen
  \bibfield  {author} {\bibinfo {author} {\bibfnamefont {P.~E.}\ \bibnamefont
  {V{\'e}rtes}}, \bibinfo {author} {\bibfnamefont {A.~F.}\ \bibnamefont
  {Alexander-Bloch}}, \bibinfo {author} {\bibfnamefont {N.}~\bibnamefont
  {Gogtay}}, \bibinfo {author} {\bibfnamefont {J.~N.}\ \bibnamefont {Giedd}},
  \bibinfo {author} {\bibfnamefont {J.~L.}\ \bibnamefont {Rapoport}}, \ and\
  \bibinfo {author} {\bibfnamefont {E.~T.}\ \bibnamefont {Bullmore}},\
  }\href@noop {} {\bibfield  {journal} {\bibinfo  {journal} {Proceedings of the
  National Academy of Sciences}\ }\textbf {\bibinfo {volume} {109}},\ \bibinfo
  {pages} {5868} (\bibinfo {year} {2012})}\BibitemShut {NoStop}%
\bibitem [{\citenamefont {Manna}\ and\ \citenamefont
  {Sen}(2002)}]{manna2002modulated}%
  \BibitemOpen
  \bibfield  {author} {\bibinfo {author} {\bibfnamefont {S.~S.}\ \bibnamefont
  {Manna}}\ and\ \bibinfo {author} {\bibfnamefont {P.}~\bibnamefont {Sen}},\
  }\href@noop {} {\bibfield  {journal} {\bibinfo  {journal} {Physical Review
  E}\ }\textbf {\bibinfo {volume} {66}},\ \bibinfo {pages} {066114} (\bibinfo
  {year} {2002})}\BibitemShut {NoStop}%
\bibitem [{\citenamefont {Xulvi-Brunet}\ and\ \citenamefont
  {Sokolov}(2002)}]{xulvi2002evolving}%
  \BibitemOpen
  \bibfield  {author} {\bibinfo {author} {\bibfnamefont {R.}~\bibnamefont
  {Xulvi-Brunet}}\ and\ \bibinfo {author} {\bibfnamefont {I.~M.}\ \bibnamefont
  {Sokolov}},\ }\href@noop {} {\bibfield  {journal} {\bibinfo  {journal}
  {Physical Review E}\ }\textbf {\bibinfo {volume} {66}},\ \bibinfo {pages}
  {026118} (\bibinfo {year} {2002})}\BibitemShut {NoStop}%
\bibitem [{\citenamefont {Boccaletti}\ \emph {et~al.}(2006)\citenamefont
  {Boccaletti}, \citenamefont {Latora}, \citenamefont {Moreno}, \citenamefont
  {Chavez},\ and\ \citenamefont {Hwang}}]{boccaletti2006complex}%
  \BibitemOpen
  \bibfield  {author} {\bibinfo {author} {\bibfnamefont {S.}~\bibnamefont
  {Boccaletti}}, \bibinfo {author} {\bibfnamefont {V.}~\bibnamefont {Latora}},
  \bibinfo {author} {\bibfnamefont {Y.}~\bibnamefont {Moreno}}, \bibinfo
  {author} {\bibfnamefont {M.}~\bibnamefont {Chavez}}, \ and\ \bibinfo {author}
  {\bibfnamefont {D.-U.}\ \bibnamefont {Hwang}},\ }\href@noop {} {\bibfield
  {journal} {\bibinfo  {journal} {Physics reports}\ }\textbf {\bibinfo {volume}
  {424}},\ \bibinfo {pages} {175} (\bibinfo {year} {2006})}\BibitemShut
  {NoStop}%
\bibitem [{\citenamefont {Poncela}\ \emph {et~al.}(2008)\citenamefont
  {Poncela}, \citenamefont {G{\'o}mez-Gardenes}, \citenamefont {Flor{\'\i}a},
  \citenamefont {S{\'a}nchez},\ and\ \citenamefont
  {Moreno}}]{poncela2008complex}%
  \BibitemOpen
  \bibfield  {author} {\bibinfo {author} {\bibfnamefont {J.}~\bibnamefont
  {Poncela}}, \bibinfo {author} {\bibfnamefont {J.}~\bibnamefont
  {G{\'o}mez-Gardenes}}, \bibinfo {author} {\bibfnamefont {L.~M.}\ \bibnamefont
  {Flor{\'\i}a}}, \bibinfo {author} {\bibfnamefont {A.}~\bibnamefont
  {S{\'a}nchez}}, \ and\ \bibinfo {author} {\bibfnamefont {Y.}~\bibnamefont
  {Moreno}},\ }\href@noop {} {\bibfield  {journal} {\bibinfo  {journal} {PLoS
  one}\ }\textbf {\bibinfo {volume} {3}},\ \bibinfo {pages} {e2449} (\bibinfo
  {year} {2008})}\BibitemShut {NoStop}%
\bibitem [{\citenamefont {Gomez-Gardenes}\ and\ \citenamefont
  {Moreno}(2004)}]{gomez2004local}%
  \BibitemOpen
  \bibfield  {author} {\bibinfo {author} {\bibfnamefont {J.}~\bibnamefont
  {Gomez-Gardenes}}\ and\ \bibinfo {author} {\bibfnamefont {Y.}~\bibnamefont
  {Moreno}},\ }\href@noop {} {\bibfield  {journal} {\bibinfo  {journal}
  {Physical Review E}\ }\textbf {\bibinfo {volume} {69}},\ \bibinfo {pages}
  {037103} (\bibinfo {year} {2004})}\BibitemShut {NoStop}%
\bibitem [{\citenamefont {Bollob{\'a}s}\ and\ \citenamefont
  {Riordan}(2003)}]{BRsurvey2003}%
  \BibitemOpen
  \bibfield  {author} {\bibinfo {author} {\bibfnamefont {B.}~\bibnamefont
  {Bollob{\'a}s}}\ and\ \bibinfo {author} {\bibfnamefont {O.}~\bibnamefont
  {Riordan}},\ }in\ \href@noop {} {\emph {\bibinfo {booktitle} {Handbook of
  Graphs and Networks}}}\ (\bibinfo  {publisher} {Wiley-VCH, Weinheim},\
  \bibinfo {year} {2003})\ pp.\ \bibinfo {pages} {1--34}\BibitemShut {NoStop}%
\bibitem [{\citenamefont {Serrano}\ \emph {et~al.}(2008)\citenamefont
  {Serrano}, \citenamefont {Krioukov},\ and\ \citenamefont
  {Bogun{\'a}}}]{serrano2008self}%
  \BibitemOpen
  \bibfield  {author} {\bibinfo {author} {\bibfnamefont {M.~A.}\ \bibnamefont
  {Serrano}}, \bibinfo {author} {\bibfnamefont {D.}~\bibnamefont {Krioukov}}, \
  and\ \bibinfo {author} {\bibfnamefont {M.}~\bibnamefont {Bogun{\'a}}},\
  }\href@noop {} {\bibfield  {journal} {\bibinfo  {journal} {Physical review
  letters}\ }\textbf {\bibinfo {volume} {100}},\ \bibinfo {pages} {078701}
  (\bibinfo {year} {2008})}\BibitemShut {NoStop}%
\bibitem [{\citenamefont {Krioukov}\ \emph {et~al.}(2009)\citenamefont
  {Krioukov}, \citenamefont {Papadopoulos}, \citenamefont {Vahdat},\ and\
  \citenamefont {Bogu{\~n}{\'a}}}]{krioukov2009curvature}%
  \BibitemOpen
  \bibfield  {author} {\bibinfo {author} {\bibfnamefont {D.}~\bibnamefont
  {Krioukov}}, \bibinfo {author} {\bibfnamefont {F.}~\bibnamefont
  {Papadopoulos}}, \bibinfo {author} {\bibfnamefont {A.}~\bibnamefont
  {Vahdat}}, \ and\ \bibinfo {author} {\bibfnamefont {M.}~\bibnamefont
  {Bogu{\~n}{\'a}}},\ }\href@noop {} {\bibfield  {journal} {\bibinfo  {journal}
  {Physical Review E}\ }\textbf {\bibinfo {volume} {80}},\ \bibinfo {pages}
  {035101} (\bibinfo {year} {2009})}\BibitemShut {NoStop}%
\bibitem [{\citenamefont {Krioukov}\ \emph {et~al.}(2010)\citenamefont
  {Krioukov}, \citenamefont {Papadopoulos}, \citenamefont {Kitsak},
  \citenamefont {Vahdat},\ and\ \citenamefont
  {Bogun{\'a}}}]{krioukov2010hyperbolic}%
  \BibitemOpen
  \bibfield  {author} {\bibinfo {author} {\bibfnamefont {D.}~\bibnamefont
  {Krioukov}}, \bibinfo {author} {\bibfnamefont {F.}~\bibnamefont
  {Papadopoulos}}, \bibinfo {author} {\bibfnamefont {M.}~\bibnamefont
  {Kitsak}}, \bibinfo {author} {\bibfnamefont {A.}~\bibnamefont {Vahdat}}, \
  and\ \bibinfo {author} {\bibfnamefont {M.}~\bibnamefont {Bogun{\'a}}},\
  }\href@noop {} {\bibfield  {journal} {\bibinfo  {journal} {Physical Review
  E}\ }\textbf {\bibinfo {volume} {82}},\ \bibinfo {pages} {036106} (\bibinfo
  {year} {2010})}\BibitemShut {NoStop}%
\bibitem [{\citenamefont {Rozenfeld}\ \emph {et~al.}(2002)\citenamefont
  {Rozenfeld}, \citenamefont {Cohen}, \citenamefont {Ben-Avraham},\ and\
  \citenamefont {Havlin}}]{rozenfeld2002scale}%
  \BibitemOpen
  \bibfield  {author} {\bibinfo {author} {\bibfnamefont {A.~F.}\ \bibnamefont
  {Rozenfeld}}, \bibinfo {author} {\bibfnamefont {R.}~\bibnamefont {Cohen}},
  \bibinfo {author} {\bibfnamefont {D.}~\bibnamefont {Ben-Avraham}}, \ and\
  \bibinfo {author} {\bibfnamefont {S.}~\bibnamefont {Havlin}},\ }\href@noop {}
  {\bibfield  {journal} {\bibinfo  {journal} {Physical Review Letters}\
  }\textbf {\bibinfo {volume} {89}},\ \bibinfo {pages} {218701} (\bibinfo
  {year} {2002})}\BibitemShut {NoStop}%
\bibitem [{\citenamefont {Warren}\ \emph {et~al.}(2002)\citenamefont {Warren},
  \citenamefont {Sander},\ and\ \citenamefont {Sokolov}}]{warren2002geography}%
  \BibitemOpen
  \bibfield  {author} {\bibinfo {author} {\bibfnamefont {C.~P.}\ \bibnamefont
  {Warren}}, \bibinfo {author} {\bibfnamefont {L.~M.}\ \bibnamefont {Sander}},
  \ and\ \bibinfo {author} {\bibfnamefont {I.~M.}\ \bibnamefont {Sokolov}},\
  }\href@noop {} {\bibfield  {journal} {\bibinfo  {journal} {Physical Review
  E}\ }\textbf {\bibinfo {volume} {66}},\ \bibinfo {pages} {056105} (\bibinfo
  {year} {2002})}\BibitemShut {NoStop}%
\bibitem [{\citenamefont {Herrmann}\ \emph {et~al.}(2003)\citenamefont
  {Herrmann}, \citenamefont {Barth{\'e}lemy},\ and\ \citenamefont
  {Provero}}]{herrmann2003connectivity}%
  \BibitemOpen
  \bibfield  {author} {\bibinfo {author} {\bibfnamefont {C.}~\bibnamefont
  {Herrmann}}, \bibinfo {author} {\bibfnamefont {M.}~\bibnamefont
  {Barth{\'e}lemy}}, \ and\ \bibinfo {author} {\bibfnamefont {P.}~\bibnamefont
  {Provero}},\ }\href@noop {} {\bibfield  {journal} {\bibinfo  {journal}
  {Physical Review E}\ }\textbf {\bibinfo {volume} {68}},\ \bibinfo {pages}
  {026128} (\bibinfo {year} {2003})}\BibitemShut {NoStop}%
\bibitem [{\citenamefont {Barab{\'a}si}\ and\ \citenamefont
  {Albert}(1999)}]{barabasi1999emergence}%
  \BibitemOpen
  \bibfield  {author} {\bibinfo {author} {\bibfnamefont {A.-L.}\ \bibnamefont
  {Barab{\'a}si}}\ and\ \bibinfo {author} {\bibfnamefont {R.}~\bibnamefont
  {Albert}},\ }\href@noop {} {\bibfield  {journal} {\bibinfo  {journal}
  {Science}\ }\textbf {\bibinfo {volume} {286}},\ \bibinfo {pages} {509}
  (\bibinfo {year} {1999})}\BibitemShut {NoStop}%
\bibitem [{\citenamefont {Bollob{\'a}s}\ and\ \citenamefont
  {Riordan}(2004)}]{BRdiam}%
  \BibitemOpen
  \bibfield  {author} {\bibinfo {author} {\bibfnamefont {B.}~\bibnamefont
  {Bollob{\'a}s}}\ and\ \bibinfo {author} {\bibfnamefont {O.}~\bibnamefont
  {Riordan}},\ }\href {\doibase 10.1007/s00493-004-0002-2} {\bibfield
  {journal} {\bibinfo  {journal} {Combinatorica}\ }\textbf {\bibinfo {volume}
  {24}},\ \bibinfo {pages} {5} (\bibinfo {year} {2004})}\BibitemShut {NoStop}%
\bibitem [{\citenamefont {Goltsev}\ \emph {et~al.}(2003)\citenamefont
  {Goltsev}, \citenamefont {Dorogovtsev},\ and\ \citenamefont
  {Mendes}}]{goltsev2003critical}%
  \BibitemOpen
  \bibfield  {author} {\bibinfo {author} {\bibfnamefont {A.}~\bibnamefont
  {Goltsev}}, \bibinfo {author} {\bibfnamefont {S.}~\bibnamefont
  {Dorogovtsev}}, \ and\ \bibinfo {author} {\bibfnamefont {J.}~\bibnamefont
  {Mendes}},\ }\href@noop {} {\bibfield  {journal} {\bibinfo  {journal}
  {Physical Review E}\ }\textbf {\bibinfo {volume} {67}},\ \bibinfo {pages}
  {026123} (\bibinfo {year} {2003})}\BibitemShut {NoStop}%
\bibitem [{\citenamefont {Dorogovtsev}\ \emph {et~al.}(2008)\citenamefont
  {Dorogovtsev}, \citenamefont {Goltsev},\ and\ \citenamefont
  {Mendes}}]{dorogovtsev2008critical}%
  \BibitemOpen
  \bibfield  {author} {\bibinfo {author} {\bibfnamefont {S.~N.}\ \bibnamefont
  {Dorogovtsev}}, \bibinfo {author} {\bibfnamefont {A.~V.}\ \bibnamefont
  {Goltsev}}, \ and\ \bibinfo {author} {\bibfnamefont {J.~F.}\ \bibnamefont
  {Mendes}},\ }\href@noop {} {\bibfield  {journal} {\bibinfo  {journal}
  {Reviews of Modern Physics}\ }\textbf {\bibinfo {volume} {80}},\ \bibinfo
  {pages} {1275} (\bibinfo {year} {2008})}\BibitemShut {NoStop}%
\bibitem [{\citenamefont {Song}\ \emph {et~al.}(2005)\citenamefont {Song},
  \citenamefont {Havlin},\ and\ \citenamefont {Makse}}]{song2005self}%
  \BibitemOpen
  \bibfield  {author} {\bibinfo {author} {\bibfnamefont {C.}~\bibnamefont
  {Song}}, \bibinfo {author} {\bibfnamefont {S.}~\bibnamefont {Havlin}}, \ and\
  \bibinfo {author} {\bibfnamefont {H.~A.}\ \bibnamefont {Makse}},\ }\href@noop
  {} {\bibfield  {journal} {\bibinfo  {journal} {Nature}\ }\textbf {\bibinfo
  {volume} {433}},\ \bibinfo {pages} {392} (\bibinfo {year}
  {2005})}\BibitemShut {NoStop}%
\bibitem [{\citenamefont {Barab{\'a}si}(2016)}]{barabasi2016network}%
  \BibitemOpen
  \bibfield  {author} {\bibinfo {author} {\bibfnamefont {A.-L.}\ \bibnamefont
  {Barab{\'a}si}},\ }\href@noop {} {\emph {\bibinfo {title} {Network
  science}}}\ (\bibinfo  {publisher} {Cambridge university press},\ \bibinfo
  {year} {2016})\ Chap.~\bibinfo {chapter} {6}\BibitemShut {NoStop}%
\bibitem [{\citenamefont {Ghoshal}\ \emph {et~al.}(2013)\citenamefont
  {Ghoshal}, \citenamefont {Chi},\ and\ \citenamefont
  {Barab{\'a}si}}]{ghoshal2013uncovering}%
  \BibitemOpen
  \bibfield  {author} {\bibinfo {author} {\bibfnamefont {G.}~\bibnamefont
  {Ghoshal}}, \bibinfo {author} {\bibfnamefont {L.}~\bibnamefont {Chi}}, \ and\
  \bibinfo {author} {\bibfnamefont {A.-L.}\ \bibnamefont {Barab{\'a}si}},\
  }\href@noop {} {\bibfield  {journal} {\bibinfo  {journal} {Scientific
  reports}\ }\textbf {\bibinfo {volume} {3}} (\bibinfo {year}
  {2013})}\BibitemShut {NoStop}%
\bibitem [{\citenamefont {Cohen}\ \emph {et~al.}(2000)\citenamefont {Cohen},
  \citenamefont {Erez}, \citenamefont {Ben-Avraham},\ and\ \citenamefont
  {Havlin}}]{cohen2000resilience}%
  \BibitemOpen
  \bibfield  {author} {\bibinfo {author} {\bibfnamefont {R.}~\bibnamefont
  {Cohen}}, \bibinfo {author} {\bibfnamefont {K.}~\bibnamefont {Erez}},
  \bibinfo {author} {\bibfnamefont {D.}~\bibnamefont {Ben-Avraham}}, \ and\
  \bibinfo {author} {\bibfnamefont {S.}~\bibnamefont {Havlin}},\ }\href@noop {}
  {\bibfield  {journal} {\bibinfo  {journal} {Physical Review Letters}\
  }\textbf {\bibinfo {volume} {85}},\ \bibinfo {pages} {4626} (\bibinfo {year}
  {2000})}\BibitemShut {NoStop}%
\bibitem [{\citenamefont {Pastor-Satorras}\ and\ \citenamefont
  {Vespignani}(2001)}]{pastor2001epidemic}%
  \BibitemOpen
  \bibfield  {author} {\bibinfo {author} {\bibfnamefont {R.}~\bibnamefont
  {Pastor-Satorras}}\ and\ \bibinfo {author} {\bibfnamefont {A.}~\bibnamefont
  {Vespignani}},\ }\href@noop {} {\bibfield  {journal} {\bibinfo  {journal}
  {Physical Review Letters}\ }\textbf {\bibinfo {volume} {86}},\ \bibinfo
  {pages} {3200} (\bibinfo {year} {2001})}\BibitemShut {NoStop}%
\bibitem [{\citenamefont {Krapivsky}\ \emph {et~al.}(2000)\citenamefont
  {Krapivsky}, \citenamefont {Redner},\ and\ \citenamefont
  {Leyvraz}}]{krapivsky2000connectivity}%
  \BibitemOpen
  \bibfield  {author} {\bibinfo {author} {\bibfnamefont {P.~L.}\ \bibnamefont
  {Krapivsky}}, \bibinfo {author} {\bibfnamefont {S.}~\bibnamefont {Redner}}, \
  and\ \bibinfo {author} {\bibfnamefont {F.}~\bibnamefont {Leyvraz}},\
  }\href@noop {} {\bibfield  {journal} {\bibinfo  {journal} {Physical Review
  Letters}\ }\textbf {\bibinfo {volume} {85}},\ \bibinfo {pages} {4629}
  (\bibinfo {year} {2000})}\BibitemShut {NoStop}%
\bibitem [{\citenamefont {Redner}(2005)}]{redner2005citation}%
  \BibitemOpen
  \bibfield  {author} {\bibinfo {author} {\bibfnamefont {S.}~\bibnamefont
  {Redner}},\ }\href@noop {} {\bibfield  {journal} {\bibinfo  {journal}
  {Physics Today}\ }\textbf {\bibinfo {volume} {58}},\ \bibinfo {pages} {49}
  (\bibinfo {year} {2005})}\BibitemShut {NoStop}%
\bibitem [{\citenamefont {Redner}(1998)}]{redner1998popular}%
  \BibitemOpen
  \bibfield  {author} {\bibinfo {author} {\bibfnamefont {S.}~\bibnamefont
  {Redner}},\ }\href@noop {} {\bibfield  {journal} {\bibinfo  {journal} {The
  European Physical Journal B -- Condensed Matter and Complex Systems}\
  }\textbf {\bibinfo {volume} {4}},\ \bibinfo {pages} {131} (\bibinfo {year}
  {1998})}\BibitemShut {NoStop}%
\bibitem [{\citenamefont {Onnela}\ \emph {et~al.}(2007)\citenamefont {Onnela},
  \citenamefont {Saram{\"a}ki}, \citenamefont {Hyv{\"o}nen}, \citenamefont
  {Szab{\'o}}, \citenamefont {Lazer}, \citenamefont {Kaski}, \citenamefont
  {Kert{\'e}sz},\ and\ \citenamefont {Barab{\'a}si}}]{onnela2007structure}%
  \BibitemOpen
  \bibfield  {author} {\bibinfo {author} {\bibfnamefont {J.-P.}\ \bibnamefont
  {Onnela}}, \bibinfo {author} {\bibfnamefont {J.}~\bibnamefont
  {Saram{\"a}ki}}, \bibinfo {author} {\bibfnamefont {J.}~\bibnamefont
  {Hyv{\"o}nen}}, \bibinfo {author} {\bibfnamefont {G.}~\bibnamefont
  {Szab{\'o}}}, \bibinfo {author} {\bibfnamefont {D.}~\bibnamefont {Lazer}},
  \bibinfo {author} {\bibfnamefont {K.}~\bibnamefont {Kaski}}, \bibinfo
  {author} {\bibfnamefont {J.}~\bibnamefont {Kert{\'e}sz}}, \ and\ \bibinfo
  {author} {\bibfnamefont {A.-L.}\ \bibnamefont {Barab{\'a}si}},\ }\href@noop
  {} {\bibfield  {journal} {\bibinfo  {journal} {Proceedings of the National
  Academy of Sciences}\ }\textbf {\bibinfo {volume} {104}},\ \bibinfo {pages}
  {7332} (\bibinfo {year} {2007})}\BibitemShut {NoStop}%
\end{thebibliography}%
\newpage
\end{document}